\documentstyle[prl,aps,twocolumn,psfig]{revtex}

\newcommand{\vect}[1]{{\mbox{\boldmath {$#1$}}}}
\newcommand{\Dt}{\frac{D}{Dt}}

\newcommand{\mev}{ {\rm MeV} }

\newcommand{\g}{ {\rm g} }
\newcommand{\cm}{ {\rm cm} }
\newcommand{\s}{ {\rm s} }
\newcommand{\erg}{ {\rm erg} }
\newcommand{\order}{{\cal O}}
\newcommand{\anti}[1]{{\overline{#1}}}

\begin{document}

\twocolumn[\hsize\textwidth\columnwidth\hsize\csname
@twocolumnfalse\endcsname

\tighten
\draft
\title{
Anisotropic $e^+ e^-$ pressure due to the QED effect in strong
magnetic fields and the application to the entropy production in
neutrino-driven wind
}

\author{Kazunori Kohri}
\address{
Yukawa Institute for Theoretical Physics, Kyoto University, 
Kyoto, 606-8502, Japan
}

\author{Shoichi Yamada}
\address{
Institute of Laser Engineering (ILE), Osaka University, Osaka
565-0871, Japan
}

\author{Shigehiro Nagataki}
\address{
Department of Physics, School of Science, the University
of Tokyo, 7-3-1 Hongo, Bunkyoku, Tokyo 113-0033, Japan
}


\maketitle

\begin{abstract}
    We study the equation of state of electron in strong magnetic
    fields which are greater than the critical value $B_c \simeq 4.4
    \times 10^{13}$ Gauss.  We find that such a strong magnetic field
    induces the anisotropic pressure of electron. We apply the result
    to the neutrino-driven wind in core-collapse supernovae and find
    that it can produce large entropy per baryon, $S \sim 400
    k_B$. This mechanism might successfully account for the production
    of the heavy nuclei with mass numbers A = 80 -- 250 through the
    r-process nucleosynthesis.
\end{abstract}

\pacs{97.60.Bw, 12.20.-m, 26.30.+k, 98.80.Ft 
\hspace{1cm} UTAP-394/01, YITP-01-52, hep-ph/0106271}


]


Recent years strongly magnetized neutron stars ($B \gtrsim 10^{15}$
Gauss) called ``magnetar'' have been reported, (see recent
compilations~\cite{gtthelf:1998,baring:2001}). Their magnitude is much
stronger than the critical value $B_c \equiv m_e^2/e \simeq 4.4 \times
10^{13}$ Gauss if we assume that their spin-down originates from the
energy loss of the electromagnetic dipole
radiation~\cite{duncan:1992}. Such a strong magnetic field is
intriguing in the astrophysical context because unlike the weak
magnetic fields, some extraordinary phenomena would be expected from
large energy splitting of Landau levels, large refractive indices of
photons, photon splitting effect and so on, (see
Refs.~\cite{Kohri:2001wx,Adler:1971wn} and references therein). In
this letter, we study the equation of state of electron in strong
magnetic fields and show that the electron pressure can become
anisotropic. So far, another papers have been published where they have
given an application to the anisotropic collapse of the neutron
gas~\cite{Chaichian:2000gd,PerezMartinez:2000jt}. We take a different
treatment in this study and show that the anisotropic pressure can
generate entropy and that it can have an implication for the r-process
nucleosynthesis in the neutrino driven wind in core-collapse
supernovae. Such a large entropy production is required for the
successful r-process nucleosynthesis in the wind to account for the
observational solar abundances~\cite{Hoffman:1996aj}.


In quantum electrodynamics (QED), the Lagrangian density of electron
is given by~\cite{Itzykson:1980rh}
\begin{eqnarray}
    \label{eq:lagrangian}
    {\cal L} =
    \overline{\psi}\left[i\hspace{-4pt}\not{\partial}-m_e\right]\psi,
\end{eqnarray}
where $\psi$ is the four-component Dirac spinor of electron, $m_e$ is
electron mass, and the slash means the contraction with Dirac's gamma
matrices $\not{p} \equiv \sum_{\mu}\gamma^{\mu} p_{\mu}$. Then, the
energy-momentum tensor density is obtained by,
\begin{eqnarray}
    \label{eq:Tmunu}
    \hat{T}^{\mu}_{ \ \nu} = \frac{\partial {\cal
    L}}{\partial(\partial_{\mu}\psi)} \partial_{\nu} \psi -
    \delta^{\mu}_{\nu} {\cal L},
\end{eqnarray}
where $\mu, \nu = 0,1,2,3$ and the metric is $\eta_{\mu \nu} = {\rm
diag}(+1, -1, - 1, -1)$. From Euler-Lagrange equation, with
Eq.~(\ref{eq:lagrangian}) we obtain the equation of motion of the free
electron, i.e., free Dirac equation, $\left[
  i\hspace{-4pt}\not{\partial} - m_e \right] \psi = 0.$
In a external magnetic field, we know the following prescription to
obtain the correct Dirac equation, $i \partial_{\mu} \to i D_{\mu} = i
\partial_{\mu} - e A_{\mu}$,
where $e$ denotes the electromagnetic coupling constant, and $A^{\mu}
= (A_0, \vect{A})$ is the vector potential. Namely we have $\left[
  i\hspace{-4pt}\not{ D} - m_e \right]\psi = 0$ in the magnetic field.
When we choose a vector potential such that $A^0 = A^1 = A^3 = 0$, and
$A^2 = Bx$, the magnetic field is $\vect{B} = (0, 0, B)$, and we
obtain the solution,
\begin{eqnarray}
    \label{eq:solutions}
    \psi = e^{-iEt}\left( \begin{array}{c} \phi \\ \chi \end{array} \right),
\end{eqnarray}
with $\phi = e^{i (p_y y + p_z z)} f_n \vect{\zeta}_{\alpha}$ and
$\chi = {\vect{\sigma} \cdot (\vect{p} - e\vect{A})}\phi /(E + m_e) $,
where $E$ is the energy of electron, $\vect{p} = (p_x, p_y, p_z)$ is
the momentum of electron, $\sigma_j$'s are Pauli matrices ($j$ = 1, 2,
3), $n$ denotes the index of Landau levels ($n = 0, 1, \cdot \cdot
\cdot$), and $\alpha$ denotes the spin index ($\alpha = 1, -1$). The
two-component spinors $\vect{\zeta}_{\alpha}$ are represented by
$\vect{\zeta}_{1}$ = $1 \choose 0$, and $\vect{\zeta}_{-1}$ = $0
\choose 1$. And, $f_n = 1/{\sqrt{2^n n!\sqrt{\pi}}} \exp[-
\xi^2/2] H_n(\xi)$,
where $\xi \equiv \sqrt{eB}\left(x-p_y/(eB)\right)$, and $H_n(\xi)$ is
a Hermite polynomial function.  Then, the electron energy is expressed
by
\begin{eqnarray}
    \label{eq:energy}
    E = \sqrt{m_e^2 + p_z^2 + eB(2 n + 1 - \alpha)}.
\end{eqnarray}
From Eq.~(\ref{eq:energy}), we see that the magnetic field breaks the
equipartition among $ p^2_x $, $(p_y-eBx)^2 $, and $p^2_z $, because
the parallel component to the magnetic field $p_z$ can have an
arbitrary value while the perpendicular components have only quantized
values as $\langle p_x^2 \rangle = \langle \left( p_y - eBx \right)^2
\rangle = eB (n + 1/2)$, where the expectation value of an operator is
defined by $ \langle \hat{\cal O} \rangle \equiv {\int \psi^{\dagger}
\hat{\cal O}\psi d^3x}/{\int \psi^{\dagger}\psi d^3x}$.

Using the above solution, we obtain each component of the
energy-momentum tensor density,
$\hat{T}^{\mu}_{\ \nu} = \overline{\psi} \left[i\gamma^{\mu} D_{\nu}
\right]\psi$.  Adopting the usual normalization in the quantum field
theory, we can integrate $\hat{T}^{\mu}_{ \ \nu}$ and obtain the
energy-momentum tensor which is defined by
\begin{eqnarray}
    \label{eq:expect}
    T^{\mu}_{ \ \nu} \equiv
    \frac{\int \hat{T}^{\mu}_{ \ \nu} d^3x} {\int {\psi}^{\dagger}\psi d^3x}.
\end{eqnarray}
It is easy to check that all the off-diagonal components vanish
completely. The diagonal components $T^{\mu}_ { \ \nu}$ = diag($E, -
\tilde{P}_x, - \tilde{P}_y, - \tilde{P}_z$) give the anisotropic
pressure which should appear in the hydrodynamic equations. They can
be read off as
\begin{eqnarray}
    \label{eq:pressure-component}
    \tilde{P}_x &=& \tilde{P}_y = (n + \frac12 - \frac{\alpha}{2})\frac{eB}{E},
    \qquad \tilde{P}_z = \frac{p_z^2}{E}.
\end{eqnarray}
Hereafter we call them ``dynamic pressures''. Compared to the case of
a weak magnetic-field limit, we easily see that only the z-component
of the dynamic pressure is unchanged. For example, $\tilde{P}_z$
becomes $E/3$ in the ultra relativistic limit. However, the
perpendicular component (x-, y-component) of the dynamic pressure
takes quantized values and is different from the one in the case of a
weak magnetic-field limit.

To discuss the dynamic pressure averaged in statistical mechanics, we
consider the grand canonical ensemble and introduce the grand
potential $\Omega$ in a unit volume in a strong magnetic filed.
\begin{eqnarray}
    \label{eq:grandpotential}
    \Omega = 
     \frac{- eB}{(2\pi)^2\beta} \sum_{\alpha = -1, 1} \sum^{\infty}_{n =
    0} \int^{\infty}_{-\infty} dp_z \ln{\left(1 + e^{-\beta(E -
      \mu)}\right)},
\end{eqnarray}
where $\beta = 1/(k_BT)$, $k_B$ is Boltzmann's constant, and $\mu$
is the chemical potential of electron. The thermodynamic pressure and
the number density of electron are obtained by $\overline{P} = -
\Omega$, and $N = - \partial \Omega/ \partial \mu$,
respectively. Then, we find that the equation of state is
$\overline{P} = N T$ even in the strong magnetic field. For the
thermodynamic pressure and the number density of positron, we can
obtain them only by changing the signature of the chemical potential
$\mu \to - \mu$ in the case of electron. Using
Eq.~(\ref{eq:pressure-component}), we can calculate the dynamic
pressure averaged in the grand canonical ensemble,
\begin{eqnarray}
\label{eq:grand-press}
P_i^ &=& \frac{eB}{(2\pi)^2} \sum^{\infty}_{l = 0} S_l
\int^{\infty}_{-\infty} dp_z \tilde{P}_i \frac1{e^{\beta(E_l - \mu)} +
1},
\end{eqnarray}
where $i$ runs $x$, $y$, and $z$, $S_l = 2 - \delta_{0 l}$, and $E_l = \sqrt{m_e^2
+ p_z^2 + 2eBl}$.

\label{sec:apply}

Next, we consider the entropy production as a consequence of the
anisotropic dynamic pressure and apply it to astrophysical conditions.
Here we discuss the r-process nucleosynthesis. It has been a mystery
for some time where and how heavy nuclei whose mass number is A = 80 --
250 are produced in the universe. Such heavy elements are supposed to
be synthesized via so-called r-process. The currently favored site is
a neutrino driven wind in core-collapse supernovae. This is because
there are a lot of free neutrons near the surface of neutron star. It
has been reported, however, that we need a very high entropy per
baryon $S \sim 400 k_B$ for the successful r-process
nucleosynthesis~\cite{Hoffman:1996aj} while both numerical simulations
and analytical treatments fall short of the requirement by a factor of
2 -- 3~\cite{Qian:1996xt}. Here we show that we can realize such a
high entropy per baryon in the strong magnetic field by way of
anisotropic dynamic pressure, and it can be a viable candidate for
successful r-process nucleosynthesis.

The first law of thermodynamics is represented by, $T \frac{DS}{Dt} =
\Dt U + \overline{P} \Dt \left[ 1/(\rho/m_N) \right]$,
where $S$ is the entropy per baryon, $T$ is the temperature, $m_N$ is
the nucleon mass, $t$ is the time, $U \equiv \epsilon m_N $ is the
internal energy per baryon ($\epsilon$ : specific internal energy),
and $\rho$ is the energy density of the matter.  $\Dt \equiv
\partial_0 + v_j \partial_j$ means the Lagrangian time derivative,
where $v_j$ is the velocity of the fluid ($j = 1, 2, 3$). Hereafter
the repeated suffix means the summation. $\overline{P}$ is the
thermodynamic pressure which is given by the grand potential as
$\overline{P} = - \Omega$. The continuity equation $\Dt \rho = \rho
\partial_j v_j$ is transformed into $\Dt \left( 1/{\rho } \right) =
1/{\rho} \partial_j v_j$.
The Euler's equation of motion, $\partial_0(\rho v_i) +
\partial_j(\rho v_j v_i) = \partial_j P_{i j}$ with the continuity
equation
is integrated as $\Dt \left( |\vect{v}|^2/2 \right) = v_i {\partial_j
P_{i j}}/{\rho}$,
where $P_{i j}$ is the stress tensor of the dynamic pressure (= $ {\rm
diag} (- P_x, - P_y, - P_Z)$ ) which corresponds to the spatial parts
of the energy-momentum tensor.  The equation of the energy
conservation, $\partial_0[\rho (|\vect{v}|^2/2 + U /m_N)] +
\partial_i[\rho(|\vect{v}|^2/2 + U /m_N)v_i - P_{i j} v_j] = 0$ is
transformed into $\Dt \left(|\vect{v}|^2/2 + {U}/{m_N} \right) =
1/{\rho} \partial_i \left( P_{i j} v_j \right)$.
%
From the above equations,
 we obtain the time-evolution equation of the
entropy,
\begin{eqnarray}
    \label{eq:dsdt}
    T \frac{DS}{Dt} = \sum^{3}_{i = 1} \frac{\overline{P} -
    P_i}{\rho/m_N} \partial_i v_i.
\end{eqnarray}

Here we emphasize that the pressure in showing up the equation of
motion is not the thermodynamic pressure $\overline{P}$ but the
ensemble averaged stress tensor $P_{i j}$. This is understood as
follows. The hydrodynamic equation can be obtained from the Boltzmann
equation by integrating over the momentum. The Boltzmann equation, on
the other hand, can be obtained from the Green function by the
so-called gradient expansion as 
\begin{eqnarray}
    \label{eq:gradient}
&&\frac{\partial f}{\partial t} + \frac{\left[p^i -
      eA^i(\vect{x}) \right]}{m_e}\frac{\partial f}{\partial x^i} +
    \frac{e\left[p^j - eA^j(\vect{x}) \right]}{m_e}\frac{\partial
    A^j(\vect{x})}{\partial x^i}\frac{\partial f}{\partial p^i}
     \nonumber \\ &&\lefteqn{= 0,}
\end{eqnarray}
in the current case, where $f$ is the distribution
function~\cite{Yamada:2000za}, (see
also~\cite{Holl:2001fs,levanda:2001}). Assuming a local thermal
equilibrium and integrating Eq.~(\ref{eq:gradient}) after multiplying
$[\vect{p} - e\vect{A}]$, we obtain the magneto-hydrodynamic equations
with the stress tensor given by Eq.~(\ref{eq:grand-press}).

To apply the above formulation to the neutrino-driven wind in
core-collapse supernovae, we adopt a simple analytic wind solution
without a magnetic field which is employed in Ref.~\cite{Qian:1996xt}
where a steady state flow and spherical symmetry are assumed. The
dynamic equations are given by
\begin{eqnarray}
    \label{eq:dynamic1}
    \dot{M} &=& 4 \pi r^2 \rho v, \\
    \label{eq:dynamic2}
    v \frac{dv}{dr} &=& -\frac1{\rho} \frac{dP}{dr} - \frac{GM}{r^2}, \\
    \label{eq:dynamic3}
    \dot{q} &=& v\left( \frac{d\epsilon}{dr} -
      \frac{P}{\rho^2}\frac{d\rho}{dr} \right), 
\end{eqnarray}
where $\dot{M}$ is the constant mass-outflow rate in the ejecta, $r$
is the radial coordinate from the center of the neutron star, $v$ is
the radial outflow velocity, $P$ is the total thermodynamic pressure,
$G$ is the Newton's gravitational constant, and $M$ is the mass of the
neutron star.  The neutrino heating rate of nucleons is represented by
$\dot{q} = \dot{q}_{\nu N} + \dot{q}_{\nu e} - \dot{q}_{eN}$, where
$\dot{q}_{\nu N}$ is a neutrino absorption rate on free nucleons ($\nu
+ N \to e + N'$), $\dot{q}_{\nu e}$ is a elastic scattering rate off
background electrons ($\nu + e \to \nu + e$), and $\dot{q}_{eN}$ is an
electron absorption rate on free nucleons ($ e + N \to \nu +
N'$)~\cite{Bethe:1993fq}.
At the temperature $T \gtrsim 0.5 \mev$, the thermal bath is
constituted of $\gamma, e^+ e^-$ and nucleons. Then, the thermodynamic
pressure and the specific internal energy are represented by $P =
11\pi^2 T^4/180 + \rho T/m_N$ and $\epsilon = 11\pi^2T^4/(60 \rho) +
3T/ (2m_N)$, respectively. On the other hand, at the temperature $T
\lesssim 0.5 \mev$ we assume $\dot{q}$=0 because electrons and
positrons disappear, and free nucleons are bound into
$\alpha$-particles or heavier nuclei.

To solve the above set of dynamic equations, we should give both the
boundary and initial conditions. Here we assume that the radius of the
neutron star $R_0$ is equal to the neutrino sphere $R_{\nu}$. Then, we
give the boundary and initial conditions there as $M = 1.4 M_{\odot}$,
$R_0 = R_{\nu} = 10$ km, and $\rho = 10^{10} \g/\cm^{3}$. The initial
velocity is chosen so that $\dot{M}$ does not become more than the
critical value $\dot{M}_{\rm crit}$ because the wind should be
subsonic, e.g., $\dot{M} = \order({10^{-6}}) M_{\odot} \ 
\erg/s$~\cite{Qian:1996xt}. We also assume that the neutrino
luminosities are identical for all the neutrino species, i.e.,
$L_{\nu} = 3.6 \times 10^{51} \erg/\s$ and take as a mean energy of
each neutrino species $\epsilon_{\nu_e} = 12 \ \mev$,
$\epsilon_{\bar{\nu}_e} = 22 \ \mev$, and $\epsilon_{\nu_{\tau}} = 34
\ \mev$. As for the initial temperature, from the analytical
treatments we estimate $T_i \simeq 3 \ \mev$ for the present
parameters~\cite{Qian:1996xt}. In addition, we fix a value of $Y_e$ to
the final one which is estimated by $Y_{e, f} \simeq 0.43$ in these
parameters~\cite{Qian:1996xt} through the whole period in which the
radial coordinate evolves from $R_0$ to the radius of the outer
boundary ($\simeq 10^4$ km) because the dynamics is not sensitive to
$Y_e$ very much.

\begin{figure}[htbp]
\vspace{-0.8cm}
\centerline{\psfig{figure=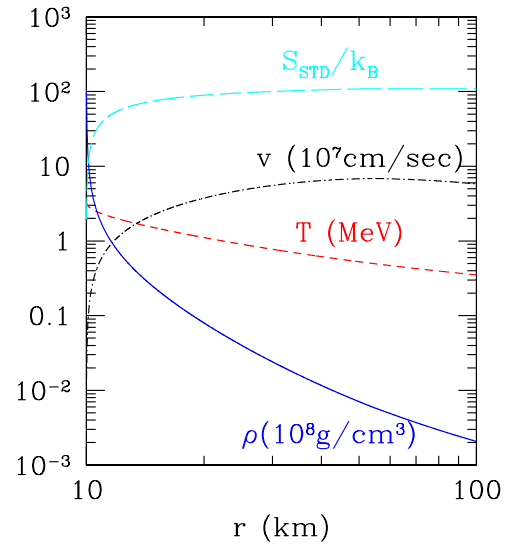,width=8.5cm}}
\vspace{-0.5cm}
\caption{
Plot of the evolution of the physical quantities as a function of the
radial coordinate $r$. They are the entropy per baryon number $S_{\rm
STD}/k_B$ (long dashed line), the radial velocity $v$ in units of
$10^7$ cm/sec (dot-dashed line), the temperature $T$ in MeV (dashed line),
and the energy density of the matter $\rho$ in $10^8 \g$/cm$^3$(solid
line).
}
\label{fig:standard}
\end{figure}

In Fig.~\ref{fig:standard}, we plot the evolutions of the physical
quantities as a function of the radial coordinate $r$. Because we are
now interested in the entropy, it is sufficient to investigate them
until $r \lesssim 100 \ {\rm km}$ which corresponds to the temperature
$T \gtrsim 0.5 \ \mev$.  Here we refer to the entropy in this model as
$S_{\rm STD}$ because this is a standard case in a weak
magnetic-field limit.

To investigate the entropy production in strong magnetic fields, we
integrate Eq.~(\ref{eq:dsdt}) using the wind solution of the above set
of dynamic equations. We see from Eq.~(\ref{eq:dsdt}) that the wind
parallel to $\vect{B}$ does not generate extra entropy. Therefore, we
consider a case in which the wind flows along the x- or y-axis and use
the radial coordinate $r$ instead of x or y. Then, we can simplify
Eq.~(\ref{eq:dsdt}) into one dimensional form and obtain the total
increment of the entropy in a strong magnetic field as
\begin{eqnarray}
    \label{eq:entpro_simp}
    \Delta S = \int \frac{P_z^e - P_x^e}{\rho T/m_N} \frac{dv}{dr} dt,
\end{eqnarray}
where the dynamic electron-positron pressure is defined by the
summation, $P_i^e = P_i^{e^-} + P_i^{e^+}$, which are presented in
Eq.~(\ref{eq:grand-press}) as $P_i^{e^-}=P_i(\mu = \mu_e)$ and
$P_i^{e^+}=P_i(\mu = - \mu_e)$. Note that $P_z^e$ is exactly equal to
$\anti{P^e}$ which is the thermodynamic pressure appearing in the
equation of state, that is, $\anti{P^e} = N_e T$ obtained from the
grand potential in Eq.~(\ref{eq:grandpotential}), where $N_e \equiv
N_{e^-} + N_{e^+}$. The number densities of electron and positron
should satisfy the following condition of the chemical equilibrium
with proton, $N_{e^-} - N_{e^+} = N_p$ by which the chemical potential
of electron $\mu_e$ is actually determined. In addition, note that
Eq.~(\ref{eq:entpro_simp}) gives the positive increment of the entropy
in the wind solution. 

\begin{figure}[htbp]
\vspace{-0.8cm}
\centerline{\psfig{figure=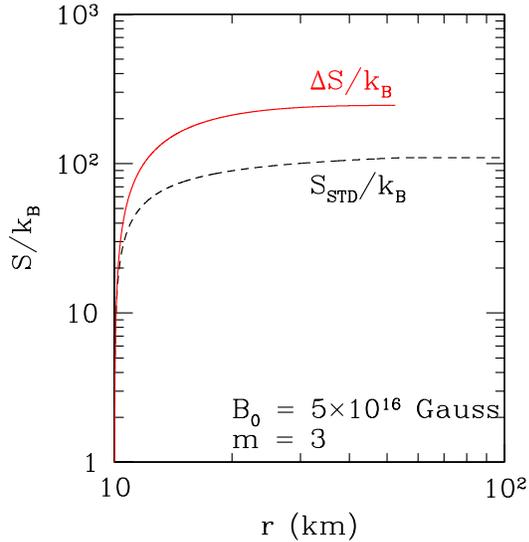,width=8.5cm}}
\vspace{-0.5cm}
\caption{
Plot of the evolution of the extra increment of the entropy in a
strong magnetic filed (solid line). For the sake of comparison, we
also plot the standard case in the weak magnetic-field limit $S_{\rm
STD}/k_B$ (dashed line).
}
\label{fig:ent}
\end{figure}

\begin{figure}[htbp]
\vspace{-.8cm}
\centerline{\psfig{figure=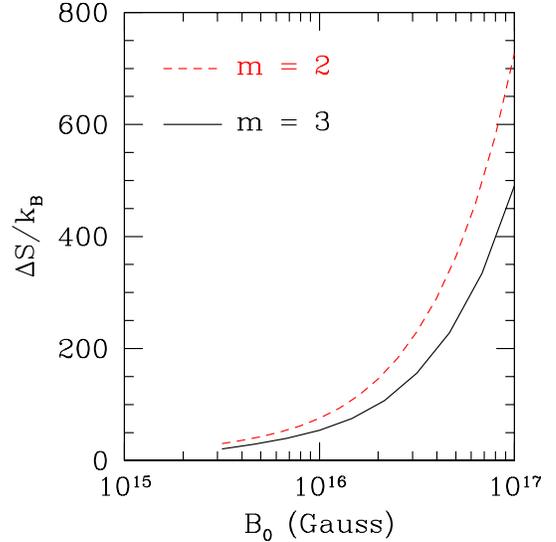,width=8.5cm}}
\vspace{-0.5cm}
\caption{
Plot of the extra entropy production $\Delta S/k_B$ as a function of
the magnitude of the magnetic field at the surface of the neutron
star. The dashed line represents the case of $m = 2$ which corresponds
to the configuration which means the flux conservation of the magnetic
field.  The solid line represents the case of $m = 3$ which
corresponds to the configuration of the dipole.
}
\label{fig:B_delent}
\end{figure}

Now we compute the cases of strong magnetic fields in which $\sqrt{eB}
\gg T \gtrsim m_e$. Then, it is expected that there exists large
anisotropy of the dynamic pressure. For simplicity, it would be
adequate for us to consider only the ground state and the next one in
the Landau levels (n = 0, 1) because of the Boltzmann suppression of
the Fermi distribution. In Fig.~\ref{fig:ent} we plot the evolution of
the entropy in a strong magnetic field. Here we assume that the
configuration of the magnetic filed is represented by $B = B_0
(r/R_0)^{-m}$, with $B_0$ the magnitude of the magnetic field at the
surface of the neutron star (= $5 \times 10^{16}$ Gauss). The index $m
= 3$ corresponds to the dipole magnetic field. From the figure we find
that the anisotropic dynamic pressure can produce the extra entropy
$\Delta S/k_B$ which is much larger than the standard value. In
Fig.~\ref{fig:B_delent} we plot the extra entropy production as a
function of the magnetic field at the surface of the neutron
star. From the figure, we can see that if the magnetic field is $B
\simeq (4-6) \times 10^{16}$ Gauss, we have a large entropy which is
required for the successful r-process nucleosynthesis.

\label{sec:conclusion}

In this study we have investigated the equation of state of electron
in strong magnetic fields $B \gtrsim B_c$ ($\simeq 4.4 \times 10^{13}$
Gauss) and found that it induces the anisotropic dynamic pressure of
electron.  We have applied the anisotropic dynamic pressure to the
r-process nucleosynthesis in core-collapse supernovae. As a result, we
obtain a large entropy $S \sim 400 k_B$ for $B \simeq (4-6) \times
10^{16}$ Gauss. This mechanism of the entropy production might
successfully give an account of the observational solar abundances of
the heavy nuclei through the r-process nucleosynthesis in the
magnetized neutrino wind.  Though in this letter we have assumed for
simplicity that the magnetic field would not influence the dynamics of
the wind, we definitely need more consistent magneto-hydrodynamic
models of the wind. That will be discussed in a separate
paper~\cite{next}.

\vspace{-0.5cm}


\end{document}